\documentclass[10pt,conference]{IEEEtran}
\usepackage{cite}
\usepackage{amsmath,amssymb,amsfonts}
\usepackage{algorithmic}
\usepackage{graphicx}
\usepackage{textcomp}
\usepackage[hyphens]{url}
\usepackage{fancyhdr}
\usepackage{hyperref}
\usepackage{soul}

\AtBeginDocument{}

\usepackage{makecell}
\usepackage{circuitikz}
\usepackage{pgfplots}
\usepackage{listings}
\usepackage{color}
\usepackage{caption}
\definecolor{mygray}{rgb}{0.98,0.98,0.98}
\definecolor{s5}{HTML}{117733}
\definecolor{a14}{HTML}{999933}
\definecolor{m4}{HTML}{332288}
\usepackage{multirow}
\usepackage{amsmath}
\usepackage{tikz}
\usetikzlibrary{shapes.geometric, positioning, arrows.meta, calc, decorations.pathreplacing, matrix}

\newcommand{\hpcayear}{2027}

\title{PG-MDP: Profile-Guided Memory Dependence Prediction for Area-Constrained Cores}

\def\hpcacameraready{} 
\newcommand\hpcaauthors{
   Luke Panayi\IEEEauthorrefmark{1}\IEEEauthorrefmark{6} \quad
   Johan Jino\IEEEauthorrefmark{1} \quad
   Sebastian S. Kim\IEEEauthorrefmark{2} \quad
   Alberto Ros\IEEEauthorrefmark{2} \\
   Alexandra Jimborean\IEEEauthorrefmark{2} \quad
   Jim Whittaker\IEEEauthorrefmark{3} \quad
   Martin Berger\IEEEauthorrefmark{4}\IEEEauthorrefmark{5} \quad
   Paul H J Kelly\IEEEauthorrefmark{1} \\
}

\vspace{4mm}

\newcommand\hpcaaffiliation{
  \IEEEauthorrefmark{1}Imperial College,
  \IEEEauthorrefmark{2}University of Murcia,
  \IEEEauthorrefmark{3}Huawei R\&D Ltd,
  \IEEEauthorrefmark{4}Montanarius Ltd,
  \IEEEauthorrefmark{5}University of Sussex
  }

\vspace{4mm}

\newcommand\hpcaemail{\IEEEauthorrefmark{6} Corresponding author: l.panayi21@imperial.ac.uk}

\author{
  \ifdefined\hpcacameraready
    \IEEEauthorblockN{\hpcaauthors{}}
      \IEEEauthorblockA{
        \hpcaaffiliation{} \\
        \hpcaemail{}
      }
  \else
    \IEEEauthorblockN{\normalsize{HPCA \hpcayear{} Submission
      \textbf{\#\hpcasubmissionnumber{}}} \\
      \IEEEauthorblockA{
        Confidential Draft \\
        Do NOT Distribute!!
      }
    }
  \fi 
}

\fancypagestyle{camerareadyfirstpage}{\fancyhead{}
  
  \fancyhead[C]{
    \ifdefined\aeopen
\else
      \ifdefined\aereviewed
\else
      \ifdefined\aereproduced
\else
\fi 
    \fi 
    \fi 
}
}

\begin{document}
\maketitle

\ifdefined\hpcacameraready 
  \thispagestyle{camerareadyfirstpage}
  \pagestyle{empty}
\else
  \thispagestyle{plain}
  \pagestyle{plain}
\fi

\newcommand{\hpcaheight}{0mm}
\ifdefined\eaopen
\renewcommand{\hpcaheight}{12mm}
\fi

\begin{abstract}

Memory Dependence Prediction (MDP) is a speculative technique to predict which stores, if any, a given load will depend on. Area-constrained cores are increasingly relevant in various applications such as energy-efficient or edge systems, and often have limited space for MDP tables. This leads to a high rate of false dependencies as memory independent loads alias with unrelated predictor entries, causing unnecessary stalls in the processor pipeline. 

The conventional way to address this problem is with greater predictor size or complexity, but this is unattractive on area-constrained cores. This paper demonstrates that targeting the \textbf{predictor working set} delivers the majority of available performance without scaling any hardware structures. We achieve this with profile-guided memory dependence prediction (PG-MDP), a hardware-software co-design to label consistently memory independent loads via their opcode and remove them from the MDP working set. These loads bypass querying the MDP and always issue as soon as possible. In the event that a labeled load incorrectly passes a store to the same address, a rollback is triggered as usual but no new MDP entry is created. Across the SPECspeed 2017 suites, PG-MDP reduces MDP load queries by 80\%, false dependencies by 84\%, and improves geomean IPC for a small (ROB=128) simulated core by 4.6\% (to within 1.2\% of the IPC when using 8x more predictor entries), with \textbf{no area cost} and no additional instruction bandwidth.

\end{abstract}

\maketitle

\section{Introduction}
\label{sec:introduction}

Memory dependence prediction (MDP) is a speculative technique used in out-of-order processors to increase available instruction-level parallelism (ILP) by predicting the stores on which a given load instruction will depend, as well as identifying loads which are memory independent.

Recent work \cite{liu24,liu26b,ssbench} exploring MDP designs in modern commercial processors finds that the storage budget allocated to memory dependence predictors is extremely small, with results suggesting even high-performance cores may not break 1KB of storage, and area-constrained efficiency-focused cores could be using as little as 50-100 bytes. For example, the Apple M-series efficiency cores measured in \cite{ssbench} are listed as having 21 fully-associative entries, with each entry containing 12 load tag bits, 12 store tag bits, 3 counter bits and 5 LRU bits, coming to just 84 bytes. This is unsurprising when considering the design constraints of these predictors; because MDPs are typically queried by both loads and stores, their tables must be highly multi-ported to ensure the pipeline is never blocked by memory operations waiting to receive predictions. For example, the open-source XiangShan processor \cite{xiangshan} implements a Store Sets-based \cite{storesets} predictor with 8 read ports. This places a high premium on area allocated to memory dependence predictors, which is further exacerbated in area-constrained cores where additional area is highly contested by performance-critical components (e.g. branch prediction), and low power usage is a top priority. As such, these smaller cores employ smaller predictors and struggle with a high rate of false positives (false dependencies) due to hash collisions/aliasing, causing loads to wait for preceding stores that are incorrectly predicted to be dependent.

\begin{figure}[!t]
    \centering
    \includegraphics[width=\columnwidth,keepaspectratio]{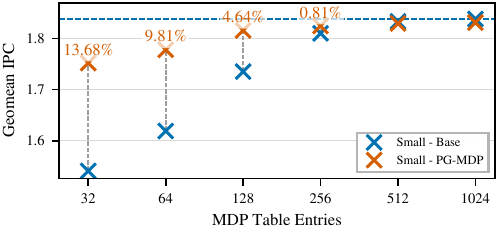}
    \caption{Base and improved IPC across SPECspeed 2017 for increasing XS Store Sets \cite{xiangshan-gem5} sizes on a small core (ROB=128). With PG-MDP, 128 entries deliver within 1.2\% of available IPC from 1024 entries.} \label{fig:ssitsweep}
\end{figure}

The importance of such area-constrained cores has also grown over time. Modern systems increasingly deploy heterogeneous processors containing both performance and efficiency-focused core designs, combining larger high-performance cores with small energy-efficient (but still out-of-order) cores, as seen in ARM's big.LITTLE systems and both Apple’s \& Intel's performance and efficiency cores. These efficiency cores share die area with high-performance cores under strict budgets for both area and power. This pattern is no longer confined to mobile devices, and this heterogeneous design can now be found across mobile, desktop, and server designs. Even as processors in high-end mobile devices scale to rival those found in laptops, they are often paired with efficiency cores to maximize battery life. Area-constrained cores are further found as standalone cores in edge systems, or efficiency tiles in chiplet architectures. In all these cases, out-of-order execution components are scaled to fit the given area envelope rather than to maximize performance. As these area-constrained cores are increasingly tasked with demanding general-purpose computation while striving to consume as little energy as possible, increasing ILP without incurring additional area or power costs becomes more pressing. 

One way to increase available ILP is to reduce false dependencies in MDP. This can be addressed by increasing predictor capacity to reduce aliasing and improve accuracy, but as established this is either unattractive or simply infeasible. Instead, this paper improves accuracy by reducing the \textbf{predictor working set}. We define the predictor working set as the set of instructions that actively index into the predictor during a given execution window. A typical MDP working set is every load and store instruction, as seen in Store Sets \cite{storesets} and many of the predictors listed in \cite{ssbench}. There exist several sophisticated hardware techniques which can at least partially reduce the MDP working set, such as load elimination \cite{constable} or value prediction \cite{vtage}, but each of these also requires space for large, power-hungry predictors, as well as abundant ILP to return the most benefit, which are precisely the resources an area-constrained core lacks. Prior work demonstrates that a considerable portion of load instructions in general-purpose workloads rarely or never exhibit in-flight dependencies even across varying inputs \cite{storedistance,arcs,constable}, and so these loads are ideal candidates to be removed from the MDP working set via hardware-software co-design in the ISA without incurring area or power cost beyond trivial additional decode logic.

We achieve this with \textbf{profile-guided memory dependence prediction (PG-MDP)}, a hardware-software co-design which identifies consistently memory-independent loads through profiling and labels them via an alternate opcode, allowing them to bypass MDP load queries and issue as soon as their register dependencies are satisfied. Labeled loads still trigger the normal squash and recovery mechanism upon a memory order violation to preserve correctness, but do not update the predictor. Using a modern variant of the Store Set predictor \cite{storesets} found in the XiangShan core \cite{xiangshan} (referred to as XS Store Sets), we show that for an appropriate working set even a small predictor is able to deliver performance competitive with a very large predictor. PG-MDP reduces average dynamic MDP load queries by 80\% and false dependencies by 84\% across SPECspeed 2017, providing a 4.6\% geomean IPC gain on a small (ROB = 128) simulated core, and within 1.2\% IPC of using an MDP with 8x more entries, as shown in Figure \ref{fig:ssitsweep}.

\subsection{Contributions}
This paper makes the following contributions:
\begin{itemize}
        \item[$\bullet$] \textbf{Reframes MDP Aliasing:} We show that the dominant source of false dependencies in memory dependence prediction does not stem solely from insufficient predictor capacity, but equally from an unnecessarily large predictor working set that captures both memory dependent and independent loads. We demonstrate that framing MDP aliasing as only a capacity problem is misleading, and that shrinking the working set is as effective as growing the predictor. 
        \item[$\bullet$] \textbf{Introduces PG-MDP:} By leveraging profile-guided memory dependence prediction, we reduce dynamic MDP load queries by 80\% on average across SPECspeed 2017, and false dependencies by 84\%.
        \item[$\bullet$] \textbf{Delivers Zero-Storage IPC Gains:} Using gem5 \cite{gem5} to simulate a core of comparable size to efficiency-focused processors found in production today, we demonstrate a 4.6\% IPC gain evaluated against a 300B (~3x expected real-world budget) XiangShan-variant Store Sets predictor (referred to as XS Store Sets) \cite{xiangshan-gem5}, without incurring any additional predictor capacity or power costs.
\end{itemize}

\section{Background}

This section outlines the fundamental concepts in scheduling memory operations in out-of-order execution. It then introduces Store Sets as a foundational memory dependence predictor algorithm as well as the modern variation found in the XiangShan processor.

\subsection{Loads in Out-of-Order Execution}
\label{sec:lsq}
A key structure in out-of-order execution is the load-store queue (LSQ), used to hold in-flight memory operations and perform memory disambiguation. When load instructions are dispatched, they are inserted into the load queue (LQ), and query the memory dependence predictor (MDP) for a predicted store(s) to wait on before execution. Once the source operands are ready and all predicted dependent stores have executed, loads search backwards through the store queue (SQ) for store-forwarding opportunities. If no matching addresses are found, loads obtain their data from the cache.

\begin{figure}[t]
    \scalebox{0.6}{\begin{tikzpicture}[
    font=\sffamily,
    >=Stealth,
    block/.style={draw, thick, rectangle, align=center, fill=white, minimum height=1cm, inner sep=5pt},
    store_entry/.style={align=left, font=\normalsize},
    load_entry/.style={align=left, font=\normalsize},
    highlight/.style={fill=orange!15}
]

\node[block, minimum width=5cm] (dispatch) at (0, 0) {\textbf{Fetch / Decode}};
\node[block, minimum width=4cm] (mdp) at (6, 0) {\textbf{Memory Dependence}\\\textbf{Predictor (MDP)}};

\draw[thick, <-] (dispatch.north) -- ++(0, 1) node[above, font=\bfseries] {New Instruction Stream};
\draw[thick, ->] (dispatch) -- (mdp) node[midway, above, font=\small] {PC};

\node[fill=gray!30, inner sep=3pt, circle] (split) at (0, -2.5) {};
\draw[thick, ->] (dispatch.south) -- (split);

\node[block, fill=blue!10, rounded corners, minimum height=0.75cm, minimum width=1.25cm] (load_split) at (2, -4.5) {\normalsize \textbf{Loads}};
\node[block, fill=blue!10, rounded corners, minimum height=0.75cm, minimum width=1.25cm] (store_split) at (2, -11.5) {\normalsize \textbf{Stores}};

\draw[thick, ->] (split) -| (load_split.north);
\draw[thick, ->] (split) -| (1, -2.5) |- (store_split.west);

\node[block, minimum width=4cm, minimum height=2cm, fill=yellow!10, align=center] (iq) at (-2.5, -8) {
    \textbf{Instruction}\\\textbf{Queue (IQ)}\\
};

\draw[thick, ->] (split) -| (iq.north) node[pos=0.75, left, font=\small]{};
\draw[line width=1.5pt, ->] (iq.south) -- ++(0,-1) node[below, font=\bfseries\Large] {\dots};

\node[fill=gray!10, inner sep=3pt, rounded corners, font=\small, draw, thick, minimum height=0.5cm, minimum width=1cm] (pred_label) at (4, -2) {Predicted Dependency};
\draw[thick, ->] (mdp.south) |- (pred_label.east);
\draw[thick, dotted, ->] (pred_label.south) |- (load_split.east);
\draw[thick, dotted, ->] (pred_label.south) |- (store_split.east);

\matrix[matrix of nodes,
        nodes={draw, thick, rectangle, minimum height=0.8cm, minimum width=4.5cm, anchor=center},
        column sep=-\pgflinewidth,
        row sep=-\pgflinewidth] (lq_mat) at (7.5, -4.5) {
    |[load_entry, highlight]| L0: $<$addr X$>$ \\
    |[load_entry]| L1: $<$addr Y$>$ \\
    |[load_entry]| L2: $<$addr Z$>$ \\
    |[load_entry]| L3: $<$addr W$>$ \\
};
\node[above=0.3cm of lq_mat, font=\bfseries\large] {Load Queue (LQ)};

\matrix[matrix of nodes,
        nodes={draw, thick, rectangle, minimum height=0.8cm, minimum width=4.5cm, anchor=center},
        column sep=-\pgflinewidth,
        row sep=-\pgflinewidth] (sq_mat) at (7.5, -11.5) {
    |[store_entry]| S0: $<$addr A$>$ \\
    |[store_entry]| S1: $<$addr B$>$ \\
    |[store_entry]| S2: $<$???$>$ \\
    |[store_entry]| S3: $<$addr X$>$ \\
    |[store_entry]| S4: $<$???$>$ \\
};
\node[above=0.3cm of sq_mat, font=\bfseries\large] {Store Queue (SQ)};

\draw[thick, ->] (load_split.east) -- (lq_mat.west |- load_split.east);
\draw[thick, ->] (store_split.east) -- (sq_mat.west |- store_split.east);

\node[left=0.3cm of lq_mat-1-1, font=\small, text=gray, align=right] (lq_head_lbl) {Oldest};
\draw[->, gray] (lq_head_lbl.east) -- (lq_mat-1-1.west);

\draw[thick, orange!80!black, ->, rounded corners=5pt]
    (lq_head_lbl.south) |- (sq_mat-4-1.west);

\node[fill=white, draw=orange!80!black, thick, font=\small, text=orange!80!black, align=center, rounded corners]
    at ($(lq_head_lbl.south) + (1.5, -3.5)$) {\textbf{Dependence Search}\\};

\node[left=0.3cm of sq_mat-1-1, font=\small, text=gray, align=right] {Oldest};
\draw[<-, gray] (sq_mat-1-1.west) -- ++(-0.3, 0);

\end{tikzpicture}
 }
    \caption{Diagram of components used to schedule memory operations in out-of-order cores. Memory operations are inserted into the instruction queue according to register dependencies and any predicted memory dependencies, and inserted into the LSQ by program order. When loads execute they search the SQ for forwarding opportunities. When stores execute they search the LQ for memory order violations.}
    \label{fig:lsq}
\end{figure}

When a load searches the SQ, not all earlier stores may have resolved their addresses. As the majority of the time a store's address will not match with the load's, it is often profitable to speculatively ignore these stores and issue the load as soon as possible. When the unresolved store eventually computes its address, it searches forward through the LQ to find loads with matching addresses which have already executed. If an address overlap is found, the load has received stale data and a memory order violation has occurred. The processor must flush all in-flight instructions from the load onwards and re-fetch.

The goal of the MDP is to minimize violation events while maximizing available ILP, allowing memory independent loads to issue as soon as possible and dependent loads to wait only for the necessary stores. When a load is incorrectly stalled on a store that does not resolve to the same address this is called a false dependency. Simple MDP algorithms typically prioritize reducing the rate of violations over the rate of false dependencies.

\subsection{Store Sets}
A seminal paper in memory dependence prediction is 'Memory Dependence Prediction using Store Sets' \cite{storesets}. This is the MDP algorithm implemented in the open source gem5 simulator \cite{gem5}. Store Sets works by using Store Set IDs (SSID) to track dependent load-store pairs, assigning each instruction the same ID value in the PC-indexed Store Set ID Table (SSIT). The SSIDs are then used to index the Last-Fetched Store Table (LFST), which holds the sequence number of the most recently issued store with an SSID mapping to that LFST entry. Newly fetched loads then access the LFST via their SSID to find the store they're predicted to be dependent on. If a load PC does not map to a valid SSID entry, it is predicted to be memory independent. To forget old dependencies and reduce table pressure, the SSIT and LFST are cyclically cleared after a certain number of fetched memory operations.

Store Sets is extremely area efficient, delivering a large portion of available performance with a tiny hardware budget. Its small size and simplicity make it well suited to area-constrained processors. However, especially at smaller sizes, Store Sets struggles with false dependencies due to hash collisions. When a memory independent load hashes to an existing SSID entry, it is incorrectly made dependent on the corresponding store for that entry. The clear period can be made shorter to offset this, but this comes at the cost of a higher rate of memory order violations as the predictor must re-learn true dependencies more often. Furthermore, many loads exhibit non-consistent memory dependencies. A load in a loop may only be dependent on a store every other iteration, and as Store Sets has no way to disambiguate these cases it conservatively predicts the load as dependent in every iteration.

\subsection{XiangShan Store Sets}
XiangShan is an open-source high-performance RISC-V processor RTL implementation \cite{xiangshan}. XiangShan represents the cutting edge in open-source superscalar processor design, and comes with a gem5 fork modified to more closely model the RTL implementation \cite{xiangshan-gem5}. This gem5 fork implements a variation of the Store Sets predictor that offers a closer representation to implementations found in real processors. 

The main optimization over original Store Sets is using multiple 'slots' for each LFST entry, thereby recording multiple dependent stores at once. This allows a load to track multiple dependencies with only one SSID. This is useful in situations where a load is dependent on different stores in different program contexts, or may have partial address overlap with multiple stores. There are further small optimizations to allocation policy which we omit here.

\section{Method}
\label{sec:method}

This section presents the profile-guided memory dependence prediction (PG-MDP) technique. We put forward the concepts of store distances, the categories of load behaviors PG-MDP labels, and the profiling algorithm to find these loads. We then motivate using profiles, and propose how load labels could be encoded in major ISAs today.

\subsection{Store Distances \& Target Load Behaviors}
\label{sec:storedistances}

This paper uses store distance to refer to the number of stores in program order between a load and its dependent store inclusive. A load depending on the most recent prior store has a store distance of 1. This concept readily maps onto processor architecture because, as overviewed in Section \ref{sec:lsq}, stores in the store queue (SQ) are also inserted in program order, so a load's store distance in program order is synonymous with the number of prior SQ entries between the load and the dependent store. Formally, let $x$ be the address of a load, and let the prior stores in program order have addresses $y_1, y_2, \dots, y_n$ ordered from youngest ($y_1$) to oldest ($y_n$). The \emph{store distance} is defined as the minimum index $i$ such that $x = y_i$. If no such index exists, the store distance is $\infty$.

The premise of PG-MDP is that loads with store distance longer than a certain threshold in 95\% or more executions are good candidates to be labeled. The optimal threshold to filter loads by is tightly tied to the hardware parameters of the target processor, so a single optimal threshold can be selected on a per-processor basis in tandem with the overall design.

The heuristic of filtering for sufficiently long store distances captures four different types of behaviors:
\begin{itemize}
        \item[$\bullet$] \textbf{(1) Memory independent loads:} Loads which read constant data (i.e. have infinite store distance).
        \item[$\bullet$] \textbf{(2) Rarely dependent loads:} Loads which observe short dependent store distances in $<$5\% of executions.
        \item[$\bullet$] \textbf{(3) Structurally independent loads:} Loads with no dependent stores in-flight at the time the load is issued.
        \item[$\bullet$] \textbf{(4) Fast-to-Resolve Dependencies:} Loads which have in-flight yet resolved dependent stores, allowing forwarding. 
\end{itemize}

Loads with behavior type (1) are always labeled regardless of the threshold value. Whether a load exhibits behaviors (2) to (4) depends on the target processor. For instance, a processor with a longer SQ will hold more in-flight stores at once, so only loads with longer store distances will be labeled. The proportion of loads that exhibit type (4) behavior is even further target specific, but is still correlated with a longer store distance as this puts more time between dispatching the store and issuing the load, making it more likely the store's address will be resolved when needed.

\subsection{Profiler Implementation \& Overhead}

For total precision this paper uses binary instrumentation on train inputs to derive store distances. The core of store distance computation is load and store memory address comparison, widely used in existing techniques such as address sanitizers \cite{asan}. The restricted scope of PG-MDP permits a simpler algorithm still. The memory address of each executed store is recorded in a FIFO queue of the same capacity as the store distance threshold of the target processor (e.g. 8). When executing a load the queue is searched to find the first prior occurrence of a matching address, if any. Each static load holds an entry in a PC-indexed hash map recording both the number of times a match did and did not occur. After program completion the hash map of load instructions is filtered for loads that did not find matches in the queue in at least 95\% of executions. The resulting list is then the load PCs to be labeled. For ease of implementation, load labels in this work are implemented in simulation using a text file of load addresses. In real deployment this list would be used to patch the program binary at the specified addresses with alternate opcodes. As instruction bandwidth is unchanged this would yield identical results.

As the compute complexity is constant for each load and store (with storage growing linearly with static loads), overhead comes almost entirely from per-instruction instrumentation itself, which could incur 10-50x slowdowns using tools such as DynamoRIO \cite{basicblockcounting}. Current trends suggest this may be tolerable, with recent work on branch prediction hardware-software co-design \cite{branchnet} proposing significantly greater overhead to chase similar performance gains. Nonetheless we will now explore possible alternatives that trade precision for overhead.

Compile-time instrumentation, used by techniques such as address sanitizers, can profile store distances at less than a 2x slowdown \cite{asan}, but is blind to stores occurring in library code. This can be mitigated by function metadata already employed by compilers to convey whether a given callsite may store to a given memory location (e.g. whether the call only touches memory passed through arguments), but beyond commonly used libraries such as libc this metadata is often incomplete and conservative (e.g. the function may store anywhere), which may limit the number of loads that can be confidently labeled.

Full binary instrumentation could be relaxed by only activating per-instruction granularity in code regions visited a certain number of times, similar to how JIT compilers determine what code to optimize. Effective hot code selection would have little to no impact on resulting performance as only cold loads would no longer be labeled. This could further be combined with statistical sampling of store distances rather than testing every execution of a load, again limiting how often per-instruction granularity is used. We aim to prototype and evaluate such techniques in future work.

\subsection{Motivating Profiles}
\label{sec:profiles}

While profile-guided optimization (PGO) has long offered performance benefits, the technique has historically struggled to achieve adoption due to complicating the compilation workflow \cite{fdocompilation}. PGO also introduces concerns about the accuracy of generated profiles, and the potential to harm performance should real input differ too much from the train input.

However, prior work shows that use of PGO has risen significantly in recent years \cite{unveilingpgo}. This can be seen in enterprise projects such as Firefox and Chrome web browsers, the Python interpreter and the GCC compiler. Performance-critical server workloads have also seen renewed attention to the benefits that PGO can provide \cite{bolt}. This supports the use of profiles as a reasonable approach to designing new hardware-software co-designs.

To show that PG-MDP generalizes, we compare which loads are labeled by profiles generated from both train and reference inputs for intspeed, shown in Figure \ref{fig:pgo}. 'Positive' means the load is labeled and 'negative' means it is not, and so true positives are loads which are labeled in both profiles, false positives are loads only labeled in the train profile, and likewise for negatives. 'Missing' means the load does not appear in the train workload. These results suggest that profiles generated from train inputs do generalize effectively to reference inputs, with the majority of loads as either true positives (TP) or true negatives (TN). Some missed potential is seen with false negatives (FN), but minimal harmful labels are introduced as false positives (FP). It is also important to note that in the case of load instructions which are unseen in the train inputs, these are never labeled and hence execute as normal. This allows PG-MDP to be more tolerant to workloads with high code coverage variation between inputs.

\begin{figure}[!t]
    \definecolor{TP}{HTML}{2E8B57}
\definecolor{TN}{HTML}{90EE90}
\definecolor{FP}{HTML}{B22222}
\definecolor{FN}{HTML}{F08080}
\definecolor{M}{HTML}{4682B4}

\begin{tikzpicture}
    \begin{axis}[
title={Ref/Train Profile Comparison},
        ybar stacked,
        ymin=0,
        ymax=100,
enlarge x limits=0.05,
        legend style={at={(0.5,-0.35)}, anchor=north, legend columns=-1},
        ylabel={\% Of Load Instructions},
        symbolic x coords={
            600.perlbench\_s,
            602.gcc\_s,
            605.mcf\_s, 620.omnetpp\_s, 623.xalancbmk\_s,
            625.x264\_s,
            631.deepsjeng\_s, 641.leela\_s, 657.xz\_s,
},
        xtick=data,
        xtick pos=bottom,
        xtick align=outside,
        x tick label style={rotate=45, anchor=east, font=\small}, yticklabel={\pgfmathprintnumber{\tick}\%},
    ]
    \legend{TP, TN, FP, FN, Missing}

\addplot+[ybar, fill=TP, draw=black] plot coordinates{
         (600.perlbench\_s,64.71)
         (602.gcc\_s,72.47)
         (605.mcf\_s,71.35)
         (620.omnetpp\_s,78.59)
         (623.xalancbmk\_s,52.24)
         (625.x264\_s,78.54)
         (631.deepsjeng\_s,70.02)
         (641.leela\_s,66.70)
(657.xz\_s,73.77)
};

\addplot+[ybar, fill=TN, draw=black] plot coordinates{
         (600.perlbench\_s,23.66)
         (602.gcc\_s,15.07)
         (605.mcf\_s,23.72)
         (620.omnetpp\_s,20.10)
         (623.xalancbmk\_s,12.65)
         (625.x264\_s,17.58)
         (631.deepsjeng\_s,27.38)
         (641.leela\_s,28.86)
         (657.xz\_s,24.50)
};

\addplot+[ybar, fill=FP, draw=black] plot coordinates{
         (600.perlbench\_s,1.19)
         (602.gcc\_s,0.77)
         (605.mcf\_s,0.47)
         (620.omnetpp\_s,0.21)
         (623.xalancbmk\_s,0.80)
         (625.x264\_s,0.59)
         (631.deepsjeng\_s,0.30)
         (641.leela\_s,1.06)
         (657.xz\_s,0.55)
};

\addplot+[ybar, fill=FN, draw=black] plot coordinates{
         (600.perlbench\_s,2.31)
         (602.gcc\_s,0.95)
         (605.mcf\_s,4.47)
         (620.omnetpp\_s,1.09)
         (623.xalancbmk\_s,0.98)
         (625.x264\_s,0.42)
         (631.deepsjeng\_s,1.45)
         (641.leela\_s,0.72)
         (657.xz\_s,1.18)
};

\addplot+[ybar, fill=M, draw=black] plot coordinates{
         (600.perlbench\_s,8.13)
         (602.gcc\_s,10.74)
         (605.mcf\_s,0.00)
         (620.omnetpp\_s,0.00)
         (623.xalancbmk\_s,33.32)
         (625.x264\_s,2.87)
         (631.deepsjeng\_s,0.85)
         (641.leela\_s,2.65)
         (657.xz\_s,0.00)
};
    \end{axis}
\end{tikzpicture}
     \caption{Comparison of labeled (static) loads between profiles generated from intspeed train and ref inputs with a store distance threshold of 8. Most labels are true positives/negatives when compared to the reference inputs, and very few are false positives.}
    \label{fig:pgo}
\end{figure}

Lastly, profiles are able to provide further benefits than static analysis. Of the four types of load behavior referred to in Section \ref{sec:storedistances}, only two (types (1) and (3)) could theoretically be captured by perfect alias analysis. Type (2) would further require perfect memory dependence analysis, and there exists no analysis in conventional compiler design that attempts to capture the behavior in type (4). It is also important to note that the alias and dependence analysis offered by industrial compilers today is far from perfect \cite{roadnottaken}. While stronger analyses do exist \cite{svf, affine}, these are either only effective in domain-specific languages or do not scale to large programs, making them either less applicable or less feasible. In contrast, profiles allow much greater flexibility and coverage of load behavior, which pairs well with speculative execution where mistakes will not invalidate program semantics.

\subsection{Encoding Labels}

PG-MDP requires only a binary ISA annotation that distinguishes conventional loads from labeled loads, and so comes with no increase in instruction bandwidth or change to binary layout. This is practically implementable in both RISC-V and AArch64, which provides coverage for the majority of efficiency-focused cores deployed today. RISC-V reserves four major custom opcodes for target-specific use \cite{riscv-custom}, and so standard $LOAD$ instructions could be mirrored to just one of these while retaining the ordinary I-type layout and other ISA semantics. In AArch64, loads are distinguished by a two-bit $opc$ field, and $opc=1x$ is left unallocated as this would encode a meaningless sign-extension \cite[\S C4-294 to C4-296]{armarm-f-c}. This leaves open an additional bit that could be used to distinguish labeled and non-labeled loads, while remaining meaningless on other targets. In both cases the ISA is able to be modified in a way where targets that benefit from PG-MDP can use labeled loads, and targets that don't see no difference. For other ISAs such as x86 where finding encoding space may not be possible, \cite{cooperative} proposes a bitmask in the program binary coupled with pre-decode logic as instructions enter the i-cache. This will incur overhead but would still maintain the critical requirement of not increasing instruction bandwidth.

\section{Evaluation}

This section presents the experimental design and results of applying PG-MDP to SPECspeed 2017. We first demonstrate the extent to which the predictor working set can be reduced. We then present a breakdown of per-workload performance gains on the small core, and show how PG-MDP performs on a medium sized core scaled to double the size of the small core. Lastly we evaluate the minimum number of MDP read ports required with and without PG-MDP to achieve maximum performance.

\subsection{Experimental Design}
\label{sec:models}

We use an optimized gem5 fork \cite{gem5-phast} that includes an implementation of the XS Store Sets predictor to model the small and medium cores. The full parameters of each model are detailed in Table \ref{table:cpu-models}. We evaluate each AArch64-compiled SPECspeed 2017 workload using up to 10 simpoints \cite{simpoints}, with an interval of 100M instructions and an additional warm-up of 10M instructions. gem5 is modified to bypass MDP queries for labeled loads and to avoid creating new entries in the case of memory order violation. Violating labeled loads still roll back as normal and do not alter program semantics.

The modeled small core is intended to resemble out-of-order efficiency cores deployed today, such as Apple's M-series efficiency cores. We stress that this correspondence is only approximate, as gem5 is a model with its own micro-architectural assumptions and features. The design class is modeled with a configuration that is appropriate to gem5 rather than by replicating any single processor. We configure XS Store Sets to use 128 SSIT entries and 64 LFST entries with two slots each, coming to a baseline-favourable 300B of MDP storage.

The medium core is used to explore how PG-MDP's impact changes as MDP size and available ILP increase, each of which is doubled over the small core. However as previously discussed this use case is somewhat less realistic as larger cores are more likely to employ additional hardware techniques which influence the baseline predictor working set.

\begin{table}[h!]
 \centering
 \caption{Parameters of processor components for each simulated model}
 \label{table:cpu-models}
 \begin{tabular}{|c|c|c|c|}
   \hline
   & & \textbf{Small} & \textbf{Med.} \\
   \hline
   \textbf{\makecell{Instruction\\ Window}} & \makecell{ROB:\\IQ:\\LQ:\\SQ:} & \makecell{128\\77\\41\\26} & \makecell{256\\154\\85\\66}\\
   \hline
   \textbf{\makecell{Pipeline\\ Width}} & \makecell{Fetch/\\Commit:\\Issue:} & \makecell{6\\8} & \makecell{8\\12}\\
   \hline
   \textbf{\makecell{XS Store Sets}} & \makecell{SSIT:\\LFST:\\Slots:\\Clear:\\} & \makecell{128\\64\\2\\125k} & \makecell{256\\128\\2\\125k}\\
   \hline
   \multirow{2}{*}{\textbf{L1D}} & \makecell{Size:\\MSHRs:} & \makecell{32KB\\16} & \makecell{64KB\\32}\\
   \cline{3-4}
   & Prefetcher: & \multicolumn{2}{c|}{Stride}\\
   \hline
   \multirow{2}{*}{\textbf{L1I}} & \makecell{Size:\\MSHRs:} & \makecell{32KB\\16} & \makecell{32KB\\32}\\
   \cline{3-4}
   & Prefetcher: & \multicolumn{2}{c|}{Tagged}\\
   \hline
   \multirow{2}{*}{\textbf{L2}} & \makecell{Size:\\MSHRs:} & \makecell{1MB\\32} & \makecell{2MB\\64}\\
   \cline{3-4}
   & Prefetcher: & \multicolumn{2}{c|}{Stride}\\
   \hline
   \multirow{2}{*}{\textbf{L3}} & \makecell{Size:\\MSHRs:} & \makecell{2MB\\64} & \makecell{4MB\\64}\\
   \cline{3-4}
   & Prefetcher: & \multicolumn{2}{c|}{Stride}\\
   \hline
   \multicolumn{2}{|c|}{\textbf{\makecell{Branch Predictor:}}} & \multicolumn{2}{c|}{\makecell{64KB TAGE-SC-L}}\\
   \hline
   \multicolumn{2}{|c|}{\textbf{\makecell{Store Distance Threshold:}}} & \makecell{8} & \makecell{22}\\
   \hline
 \end{tabular}

\end{table}

\subsection{Performance}
\label{sec:ipc}

\begin{figure*}[t]
  \centering
  \includegraphics[width=2\columnwidth, keepaspectratio]{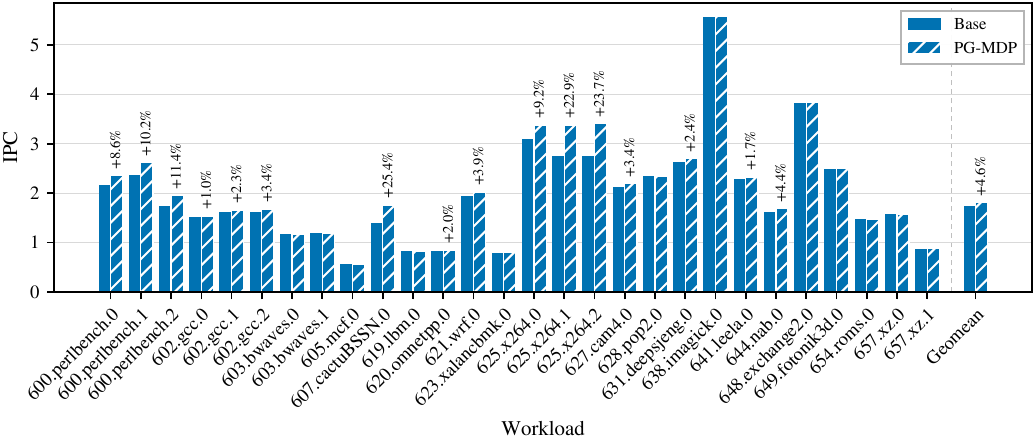}
\caption{IPC \% improvement with PG-MDP for each workload on the small core configuration. A handful of workloads benefit significantly, whereas others are largely unchanged. Annotations omit changes $<$ 0.5\%.}
\label{fig:ipc}
\end{figure*}

\begin{figure*}[t]
  \centering
  \includegraphics[width=2\columnwidth, keepaspectratio]{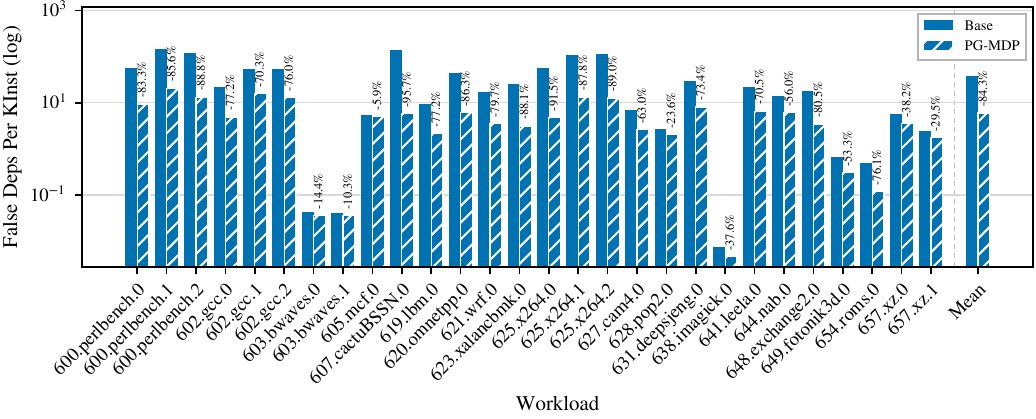}
\caption{Percent change of false dependencies per kilo-instruction. PG-MDP benefits workloads most that by default have a higher than average rate of false dependencies, but is still able to cut false dependencies across all workloads.}
\label{fig:falsedeps}
\end{figure*}

\begin{figure*}[t]
  \centering
  \includegraphics[width=2\columnwidth, keepaspectratio]{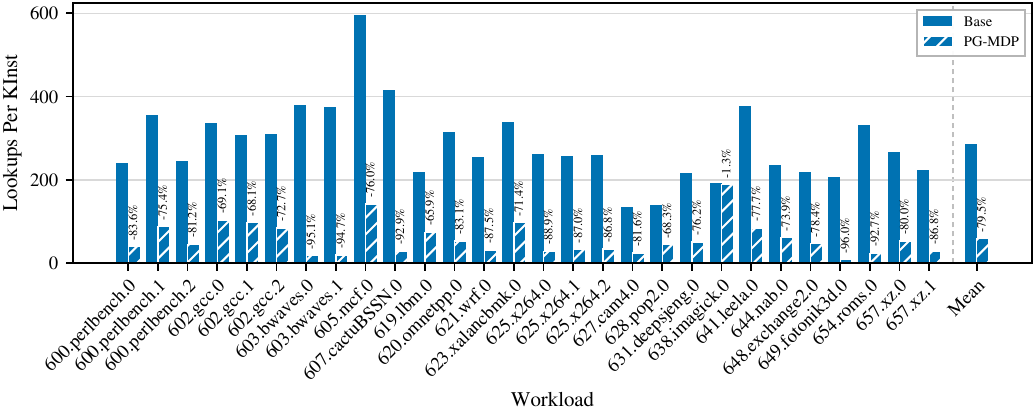}
\caption{Percent change of MDP load queries per kilo-instruction. Demonstrates the extent to which PG-MDP reduces the MDP working set by reducing how often the MDP is queried for predictions.}
\label{fig:lookups}
\end{figure*}

\begin{figure*}[t]
  \centering
  \includegraphics[width=2\columnwidth,keepaspectratio]{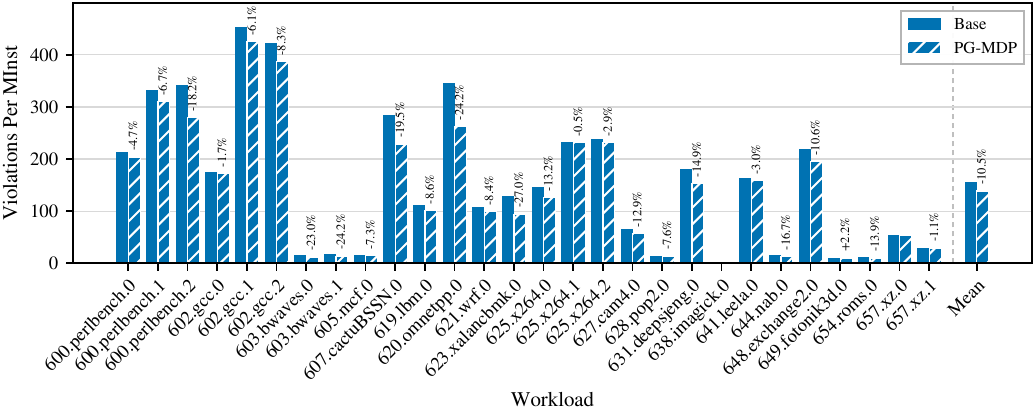}
\caption{Percent change of violations per mega-instruction. Theoretically violations may increase due to behavior differences in the train inputs, but we instead find they decrease across workloads. Annotations omit changes $<$ 0.5\%.}
\label{fig:violations}
\end{figure*}

Figure \ref{fig:ipc} shows the percent IPC changes in each SPECspeed workload for the small core configuration. It can be seen that IPC gains have an uneven spread across workloads, with some seeing large benefits (up to 25\%) and others next to none. This correlates with the magnitude of false dependencies before and after applying PG-MDP seen in Figure \ref{fig:falsedeps}. The largest gainers such as $607.cactuBSSN\_s$ and $625.x264\_s$ all exhibit relatively higher false dependencies per kilo-instruction in the base case, and see this cut down significantly by over 80\% with PG-MDP. Figure \ref{fig:lookups} further shows the reduction in dynamic MDP load queries per kilo-instruction, demonstrating the impact on predictor working set. Figure \ref{fig:violations} shows the change in memory order violations.

Listing \ref{fig:x264} uses an example function from $625.x264\_s$ to demonstrate why PG-MDP is impactful. The code is a simplified version of $pixel\_avg$, a hot function that causes a large portion of false dependencies. The loads on $src1$ and $src2$ never observe memory dependencies during program execution. As the loop is both short in instructions and part of a tight nest, these loads easily fall on the critical path during execution. Due to hash collisions, the loads are falsely made dependent on unrelated stores and the loop is unnecessarily serialized. Furthermore, the loop is compiled into three distinct variants that load different data sizes depending on remaining pixels, increasing both static load and store count pressure and making hash collisions more likely. Using PG-MDP, loads from all iterations across all access sizes are free to issue in parallel, significantly improving ILP. Altogether this represents a perfect use case for PG-MDP and is a significant contributor to the resulting performance gain. 

\begin{lstlisting}[language=C,  label=fig:x264]
void pixel_avg( uint8_t *dst, uint8_t *src1,  uint8_t *src2, 
              int i_width, int i_height )
{
    for ( int y = 0 .. i_height )
    {
        for( int x = 0 .. i_width )
            dst[x] = ( src1[x] + src2[x] + 1 ) >> 1;
    }
}
\end{lstlisting}

\begin{figure*}[t]
  \centering
  \includegraphics[width=2\columnwidth, keepaspectratio]{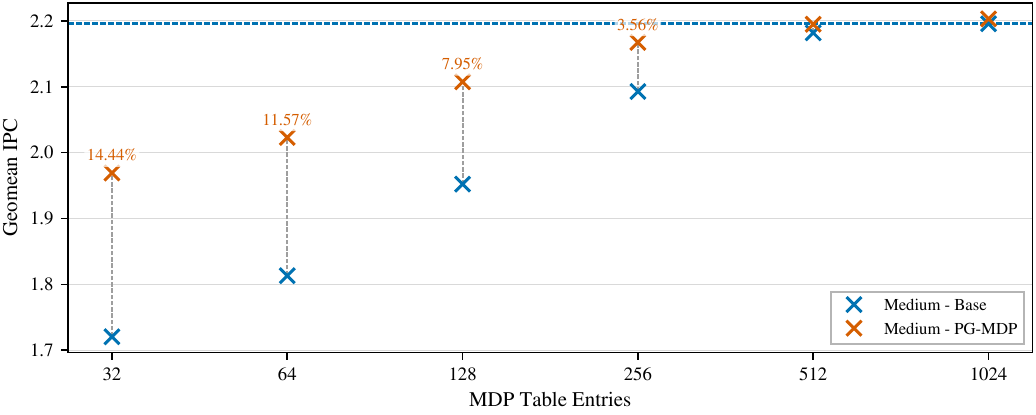}
  \caption{Base and improved IPC across SPECspeed 2017 for increasing XS Store Sets \cite{xiangshan-gem5} sizes on a medium core (ROB=256). IPC gains increase over the small core, but for the baseline MDP size (256 entries), IPC improvement is still 1\% smaller.} 
\label{fig:ipc_medium}
\end{figure*}

Figure \ref{fig:ipc_medium} shows the same MDP size sweep with and without PG-MDP evaluated on the medium core. As expected, the greater available ILP increases PG-MDP's impact on performance. However this is out-competed by using a larger MDP which diminishes opportunities for PG-MDP to reduce false dependencies, reducing the total IPC gain from 4.6\% to 3.6\%. As such while PG-MDP is still able to deliver gains to larger cores, these results confirm that the technique is best suited to smaller cores that cannot afford larger MDPs.

\subsection{Reducing MDP Read Ports}

Because PG-MDP reduces the MDP working set so significantly, it could permit designing MDPs with fewer read ports. To test this, we extend gem5 to model MDP read ports and evaluate how many ports are required to achieve maximum performance with and without PG-MDP. This is implemented by constraining the maximum number of both loads and stores which can be dispatched in a cycle before the pipeline stage blocks. We optimistically assume MDP queries always return within the current cycle, meaning this counter resets to zero every cycle. For example, 8 read ports mean any combination of 8 load or store instructions may be dispatched in a cycle, and encountering more will stall dispatch until the next cycle. Loads labeled by PG-MDP do not occupy a read port and so do not increment the counter.

\begin{figure*}[t]
  \centering
  \includegraphics[width=2\columnwidth,keepaspectratio]{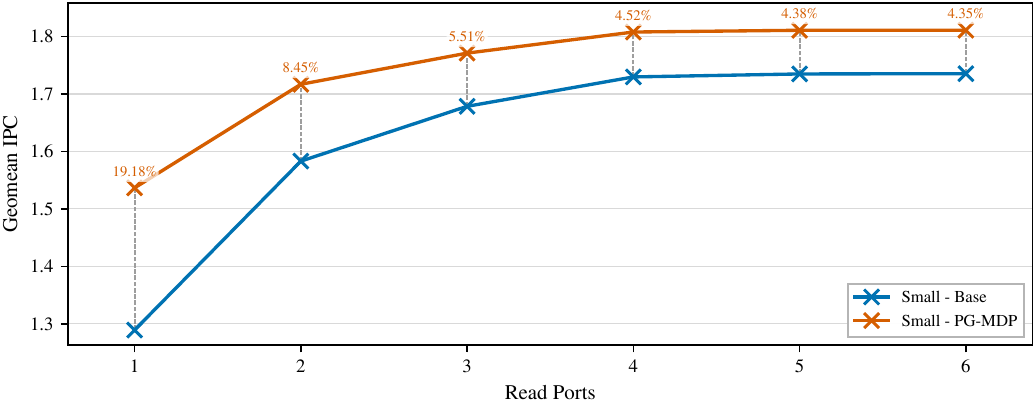}
  \caption{Geomean IPC of the small core across SPECspeed 2017 against the number of MDP read ports, with and without PG-MDP. Using PG-MDP requires three fewer ports to achieve within 1\% of using five in the baseline.}
\label{fig:readports}
\end{figure*}

Figure \ref{fig:readports} shows the geomean IPC of the small core with and without PG-MDP against the number of MDP read ports. In the baseline, 5 read ports are required for maximum performance, whereas PG-MDP sees higher IPC with only 3, and comes within 1\% of the maximum baseline performance with only 2. This suggests shrinking the working set can help overcome this design constraint in MDP, and could also make growing the predictor more feasible.

This evaluation should not be confused with a claim of how many MDP read ports real processors require. We do not evaluate predictor latency, nor the low level techniques that may or may not hide that latency, and so these results must be understood in a relative sense, i.e. through the comparison between baseline and PG-MDP, instead of as an absolute claim. 

\section{Related Work}

\emph{Improving Memory Dependence Prediction with Static Analysis} \cite{arcs} tackles the same problem as this paper, but by using static analysis to determine which load instructions to label. It demonstrates small ($<$1\%) IPC gains in select workloads, but with an overall best-case geomean improvement of less than 0.1\%. This is due to only reducing average false dependencies by 7.5\%, over 10x less than PG-MDP. Furthermore, due to relying on static analysis it also sees notable performance regressions elsewhere, hurting viability. In comparison, by leveraging profiles we are able to achieve much higher IPC gains while also avoiding significant regressions. 

\emph{Feedback-directed Memory Disambiguation Through Store Distance Analysis} \cite{storedistance} proposes a similar profile-guided co-design to this paper, but with a more aggressive use case. In their work, the MDP is replaced altogether and load instructions are extended to include 4-bit distance fields indicating the store queue position they're likely to be dependent on. This achieves high accuracy compared to Store Sets, but comes at the cost of higher instruction bandwidth to encode predicted store distances, which is usually infeasible in modern processor designs where instruction bandwidth is critical. Larger processors would also require more bits to encode longer store queue distances. Furthermore, \cite{storedistance} is more sensitive to profile accuracy than our work as loads which do not appear in the profile must be assumed to be memory independent, likely leading to a spike in memory order violations in workloads like $623.xalancbmk\_s$ with low common code coverage between train and ref inputs. In contrast, our technique is able to leave unseen loads to execute as normal. Attempting to replace the MDP altogether also hurts deployment feasibility, as it requires all binaries be compiled with this profile-guided technique, whereas our technique only supplements the MDP and so allows unlabeled binaries to still execute at baseline performance.

\emph{Software-hardware Cooperative Memory Disambiguation} \cite{cooperative} is a binary analysis co-design to reduce load queue pressure. Certain kinds of provably memory independent loads are labeled and prevented from being inserted into the load queue when issued. This technique is able to label 40\% of static loads across SPEC2006 floating point (but dynamic percentage is unclear). This also incurs additional instruction bandwidth by inserting barrier-like instructions when entering loops to ensure labeled loads are only removed from the load queue when their parent loop reaches a steady state in the pipeline. LSQ entries must also store an additional bit to track whether these custom instructions are in-flight or not. This work presents an interesting proposal for reducing load queue pressure in area-constrained processors that lack space for hardware-based load elimination techniques, but has reduced viability due to lacking a mechanism for ensuring memory coherence in multi-core systems.

\emph{Constable: Improving Performance and Power Efficiency by Safely Eliminating Load Instruction Execution} \cite{constable} is a pure hardware technique to track and eliminate stable loads which consistently read the same value from memory, i.e. behavior type (1) specified in Section \ref{sec:storedistances}. This technique is able to eliminate both the memory read and address calculation, greatly improving pipeline efficiency. It also demonstrates a broad coverage across a variety of workloads, improving geomean IPC by 5.1\% on a high-performance core. For high-performance cores this technique would be effective for reducing the MDP working set, however as discussed in Section \ref{sec:introduction}, expensive hardware techniques are less viable on area-constrained cores due to area \& power demands, as well as requiring abundant ILP to effectively return on investment.

\emph{Load-Wait Tables} A load-wait table is a very simple type of memory dependence predictor found in the Alpha 21264 \cite{kessler1999alpha}. It consists of a PC-indexed bitvector which tracks whether a given load has ever triggered a memory order violation, stalling it on all prior stores on hit and issuing as soon as possible on miss. This aims to answer the same question in hardware that PG-MDP does in software, i.e. is a given load memory independent. One might imagine a combined approach of filtering the working set of a more sophisticated predictor with a small load-wait table. We argue that PG-MDP still compares favorably against such a scheme; a load-wait table would still be queried by all loads, and so inherits the same multi-porting constraint discussed in Section \ref{sec:introduction}. It would then also suffer from table saturation at any reasonable hardware budget, and so would need to be periodically cleared and re-trained to adapt to program phase changes. This would cause spikes in false dependencies during saturation and spikes in memory order violations after clearing, limiting the accuracy it can achieve despite still incurring additional hardware costs.

\emph{Other Memory Dependence Prediction Algorithms} Besides Store Sets and load-wait tables \cite{storesets,kessler1999alpha}, other MDP algorithms include Store Vectors, MDP-TAGE, and PHAST/MASCOT \cite{storevectors, mdp-tage, mdp-tage2, phast, Mose25} (MASCOT is an extension of PHAST, but with a focus on speculative-memory bypassing and performs similarly for strictly MDP). Each of these algorithms offer different trade-offs and advantages. We choose to evaluate a Store Sets-based predictor because it achieves the highest performance of (feasible) predictors in the literature for small cores \cite{mdp-tage2}. We chose XS Store Sets \cite{xiangshan-gem5} specifically as it is, to the best of our knowledge, the most modern implementation of Store Sets and closest available representation of MDP algorithms found in real processors.

\section{Future Work}
\label{sec:futurework}

\emph{Per-Workload Threshold Selection} could achieve higher performance than per-core thresholds. As the threshold is untied to any architectural state, it is possible to select a different optimal store distance threshold for each individual workload. We chose not to do this in this work to prove that meaningful performance gains were possible with a generic threshold. However, there are use cases where this is feasible and it would be worthwhile to evaluate the potential performance available for the additional offline cost.

\emph{Just-in-Time Compilers (JITs)} are a widely used run-time based compilation technique for dynamic languages. PG-MDP could pair well with these compilers for various reasons. Firstly, the additional workflow complexity of profiling can be automated by the run-time optimizer. Secondly, JITs make speculative optimizations about observed program behavior, and so could aggressively apply ISA labels to any loads which read from addresses not recently stored to. Lastly, dynamic languages tend to be implemented with many memory independent pointer-chasing loads. This suggests these workloads could have a high potential for performance gains from using PG-MDP.

\emph{Serverless Workloads} are characterized as a collection of many short-lived workloads which only execute a few times, then not again for a long time with lots of unrelated code in-between. This effectively makes their execution always cold on the target processor, so high performance is difficult to achieve. When memory dependence predictors cannot learn dependencies fast enough to deliver performance gains, it may be preferable to disable memory dependence prediction entirely and conservatively stall loads on any unresolved stores. In this case, PG-MDP could potentially deliver substantial IPC gains, even on large cores, by allowing likely-independent loads to still issue immediately.

\section{Conclusion}
This paper proposed profile-guided memory dependence prediction (PG-MDP), a profile-guided co-design to label load instructions via their opcode and prevent them from making MDP queries. This was shown to reduce the predictor working set and reduce false dependencies by 84\%, yielding a 4.6\% IPC gain on an area-constrained core, achieving performance within 1.2\% of using 8x more predictor entries. These results point to a promising new direction for improving IPC on area-constrained, efficiency-focused cores; by attacking predictor working set size in software, the penalties of small capacity hardware structures can be almost entirely mitigated while staying within the strict design constraints of these cores.

\bibliographystyle{IEEEtranS}
\bibliography{bibtex}

\end{document}